\documentclass[aps,twocolumn,floats,nofootinbib]{revtex4}
\usepackage{graphics,graphicx,epsfig}
\usepackage{amssymb,color}
\usepackage{epsf,epstopdf,wrapfig}
\usepackage {amsmath}

\begin{document}

\title{Keith Brueckner (1924--2014).  A biographical memoir}

\author{William Bialek}
\affiliation{Joseph Henry Laboratories of Physics, Princeton University, Princeton NJ 08544 USA}
\date{\today}

\begin{abstract}
Keith Brueckner was a theoretical physicist of considerable technical power who came of age as the mysteries of the atomic nucleus were coming into focus.  His fundamental contributions to the ``many-body problem'' had a lasting impact on our understanding of how the macroscopic behavior of matter  emerges from the underlying microscopic rules.   A passionate and accomplished mountain climber, he  listed the American Alpine Club below the National Academy of Sciences on his vitae.  During decades of complex interactions between the physics community and the United States government, he helped build structures that allowed him and many others to provide advice on classified matters, but also actively raised funds to support opposition to the war in Vietnam.  At the peak of his career, he left  the Ivy League to help found and build a new university in  a small village filled with Marines and retirees---La Jolla, California.  
\end{abstract}

\maketitle

\section{Introduction}

Keith Allen Brueckner was born 19 March 1924 in Minneapolis, Minnesota.  His father Leo was Professor of Education at the University of Minnesota,  an author of mathematics textbooks, and an adviser on educational policy.  His mother Agnes (n\'ee Holland) would take a very active role in Keith's university education during World War II.  Some combination of nature and nurture produced an intensity and drive in all four of their children.  Keith's twin brother John was a gifted linguist, wrote a French contextuary for students, and taught high school;  his older brother Richard became an insurance executive but also worked as an attorney on free speech cases;  and his younger sister Patricia became a poet.

Keith attended public schools in Minneapolis and went on to the University Minnesota in 1941.  His first degree was based on a combination of course work at the university and extension courses during his military service.  His assignment was as a weatherman in the Caribbean, where his mother  sent a steady stream of the ``great books.''    While perhaps not the most dramatic thing to be doing during World War II, Keith took pride in his service.  Those who only knew the gruff and intimidating senior scientist might have been surprised to hear him break into song:
\begin{quote}
	\noindent We are the men, the weather men\\
	We may be wrong, oh now and then\\
	But when you see, those planes on high\\
	Just remember, \\
	\null \hskip 0.2 in we're the ones who let them fly
\end{quote}
After the war, Keith returned to the University of Minnesota for a year, collecting an MA, and then moved to the University of California at Berkeley for his PhD.  The 184 inch cyclotron had started running at full energy shortly before his arrival, and Berkeley was the center of an exciting interplay between theory and experiment as prewar nuclear physics evolved into postwar particle physics.  Keith tried his hand at experiments, and then found his calling as a theorist, using very general arguments to understand the recent discovery that bombarding a nucleus with X--rays could produce the elementary particles called mesons.   His PhD adviser was Robert Serber and his first theoretical paper was written with Marvin (Murph) Goldberger; Keith and Murph would remain friends for life.

\begin{figure}[b]
\includegraphics[width=\linewidth]{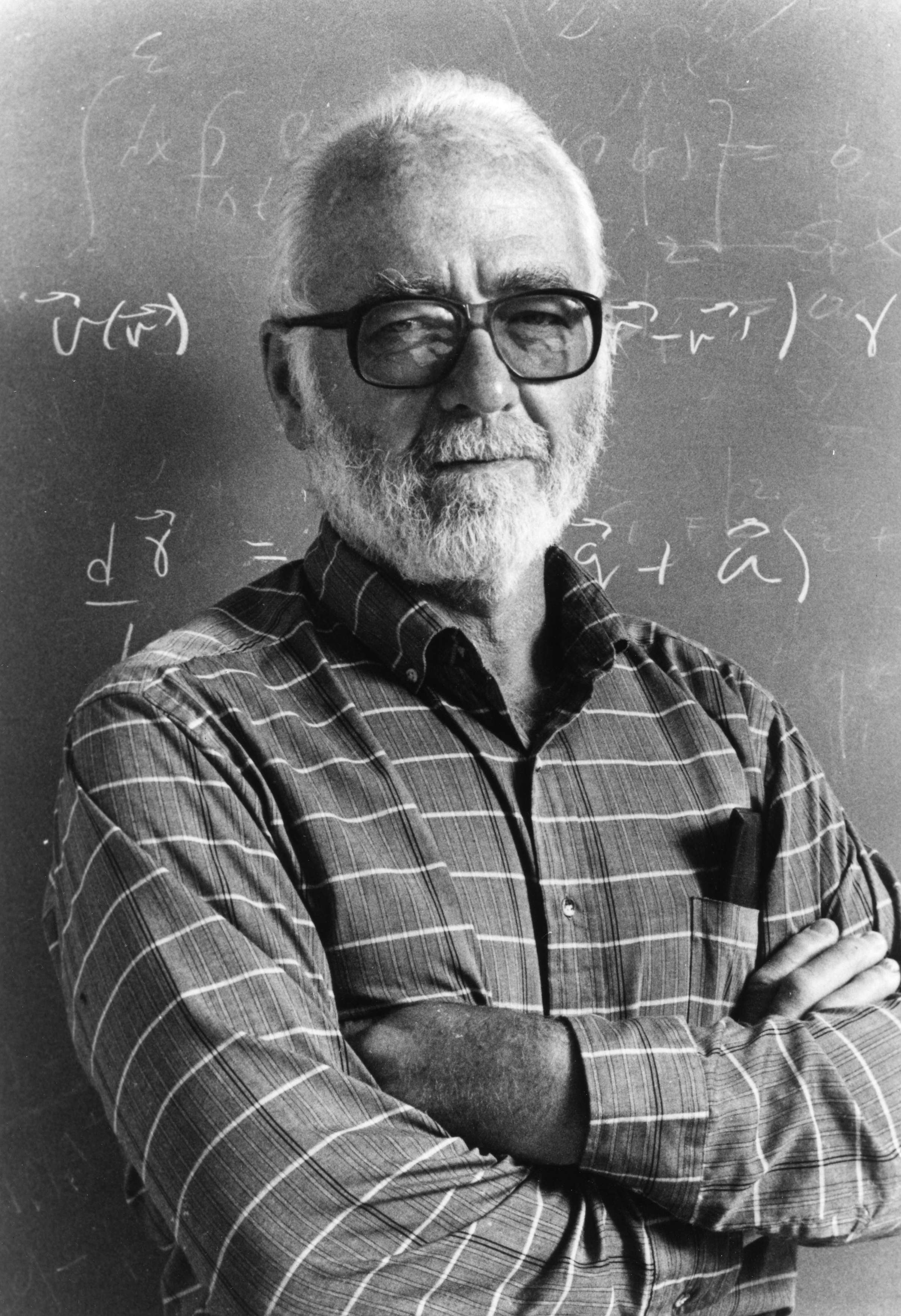}
\caption{Keith Brueckner, circa 1980.  From the American Institute of Physics Emilio Segr\`e Visual Archives.}
\end{figure}

In 1950, PhD in hand, Keith went east to the Institute for Advanced Study (IAS).  Given the long postdoctoral orbits into which our students often are launched today, it seems remarkable that he spent just a year there before becoming an Assistant Professor of Physics at Indiana University.    During that year he overlapped with Freeman Dyson, Murray Gell--Mann, Francis Low, and others with whom his life and career would intersect many times. The summer between the IAS and Indiana was spent at the University of Illinois, where he and  Gell--Mann contributed to von Neumann's classic work on reliable computation with unreliable components.

Keith's first major work at Indiana was about the interpretation of meson--nucleon scattering data, leading to the conclusion that nucleons had an excited state with quantum numbers that could be extracted from experiment.  This was near the beginning of the explosion of evidence for ``resonances,''  quickly followed by discovery of many new strongly interacting particles.  Soon Keith's attention would turn from the internal dynamics of single  particles to the behaviors that emerge when many particles interact.

\section{Many--body theory}

In a box with $N=100$ particles, there are $N^2 = 10,000$ ways for them to interact in pairs; if we consider triplets there are $N^3 = 1,000,000$  possibilities, and so on. Nonetheless, the energy comes out---experimentally---to be proportional simply to the number of particles $N$.  This happens with the molecules in the air around us, and it happens in the droplet of protons and neutrons (nucleons) that comprise a large nucleus, where the interactions among particles are vastly stronger. The surprising  fact that the energy and volume of ``nuclear matter'' are proportional to the number of particles was described in the jargon of the time as the saturation of nuclear forces.  If one tried to calculate the properties of nuclear matter using the methods available circa 1950, one found a series of terms with the factors of $N^2$, $N^3$, $\cdots$.  Keith liked to say that when he was young, problems were simple: there is a well defined calculation that should work, but gives nonsense.

Keith  showed  that the  series could be organized so that all offensive terms cancel, and the only terms left are proportional to $N$.  Keith described these surviving terms  as  linked clusters;  those which cancelled were unlinked.  Jeffrey Goldstone soon showed that this was  true to all orders of perturbation theory.  Further, one could adapt the diagrams that Richard Feynman had introduced to organize calculations in quantum field theory, and now ``linked'' and ``unlinked'' referred literally to the structure of these diagrams.

I once asked Keith what he had missed that left room for Goldstone.  He responded that he had ``just'' checked the patterns out to fourth order, and since it worked he stopped.  He would bring this raw calculational power to bear on problem after problem throughout his career.

Despite the very strong interactions among nucleons, Maria Goeppert Mayer and Hans Jensen had shown  that one could understand many properties of nuclei in a ``shell model'' where the constituent particles moved as if they were nearly independent.  This was even more surprising than the saturation of nuclear forces.  In a series of papers, Keith and his colleagues showed that the strong but  short--ranged nature of nuclear forces meant that the linked cluster expansion was dominated (in modern terms) by a particular class of ``ladder'' diagrams, and that one could sum an infinite series of these terms.  This summation embodied the intuition that when two particles collided, they would interact many times before escaping from one another.   As a  result,   single nucleons move almost independently through a ``self--consistent field'' created by the other particles.  These calculations, in collaboration with John Gammel,  would become very complex, taxing the most powerful computers of the time, but they gave semi--quantitative agreement with the extrapolated properties of nuclear matter.

Keith's work on nuclear matter was far from the last word on the subject.  It seems to have been difficult, at the time, to disentangle approximations that had physical content from those which were made for numerical convenience.  Hans Bethe played a crucial role in clarifying and reformulating what Keith had done, and the two maintained a vigorous correspondence, sharing their enthusiasm for the details of complex calculations.  A large community continues to build on the foundation that Keith laid in the 1950s, and the nuclear many--body problem has renewed relevance to the dynamics of nucleosynthesis, neutron stars, and supernovae.

In contrast to the interactions among protons and neutrons in nuclear matter, the Coulomb interactions between electrons in a metal are relatively weak, but they extend over very long distances.  In effect every electron can ``reach'' every other electron, and the total interaction energy is in danger of becoming infinite.  In the early 1950s a variety of intuitive approaches were introduced to tame this infinity.   Armed with the linked cluster expansion, Keith collaborated with Murray Gell--Mann to show that the divergences were dominated by a class of ``ring'' diagrams, and they summed an infinite set of these terms to recover finite answers, systematizing the intuitive arguments.\footnote{Keith eventually returned to this problem with his young colleague Shang--Keng Ma, studying the energy of the electron gas with spatially varying density.}  As students, twenty years later, this still seemed like magic to many of us.   

A snapshot of Keith's thinking about  these issues can be found in his lectures at one of the first of the famed Les Houches Summer Schools on Theoretical Physics.  His lectures, and those of his colleagues from both sides of the Iron Curtain,  captured the excitement of realizing that the electron gas, nuclear matter, superconductivity, and superfluidity really were all part of a single subject. Les Houches, looking across to Mont Blanc, also connected with Keith's interest in mountaineering, and he would return to the region many times.  

Keith traveled to Moscow in May 1956 for a conference hosted by the Academy of Sciences of the USSR.  He was in a group of thirteen US physicists,\footnote{Luis Alvarez,
Keith Brueckner, 
Owen Chamberlin, 
Murray Gell--Mann, 
Robert Marshak,
John Marshall, 
Abraham Pais, 
Wolfgang Panofsky, 
Emilio Segr\`e, 
LW Smith,  
Jack Steinberger, 
Victor Weisskopf,  and
Robert Wilson.  See {\em National Science Foundation 6th Annual Report for the Fiscal Year Ended June 30, 1956} (US Government Printing Office, Washington DC, 1957).} and it was a remarkable moment.  The Cold War was in full force, the Warsaw Pact had been formed the year before, and Nikita Khrushchev had spoken to a  closed session of the Communist Party Congress near the end of February 1956, beginning the process of ``de--Stalinization;'' news of this upheaval would spread slowly, appearing in the {\em New York Times} only in early June.\footnote{During the conference, Stalin was still in the mausoleum next to Lenin, but in crucial rooms his portrait had been replaced.  See LW Alvarez, Further excerpts from a Russian diary, {\em Physics Today} {\bf 10,} 22--32 (1957). } More importantly for our story, Lev Landau submitted ``The theory of a Fermi liquid'' to the  Journal of Experimental and Theoretical Physics (USSR) in March 1956, but it would not appear in English translation until January 1957.

Keith enjoyed telling the story of his meeting with Landau, a legendary figure who  gave him quite a hard time.  Why, Landau asked, were you trying to calculate properties of the ground state of many--body systems, starting from the microscopic interactions?  Wouldn't it be easier and more productive to focus on the lowest energy excitations of the system above the ground state?  To his credit, Keith admitted that he simply didn't understand what Landau was saying---until January 1957, when he could read the Fermi liquid paper.

Keith's work on the many--body problem  gave very detailed examples of how relatively simple behaviors could emerge from the daunting complexity of interacting systems.  Landau, among other insights, saw (without detailed calculation) that this simplicity allows powerful predictions to be made about excitations, such as the way heat is conducted through a fluid or electrical current through a metal.  These ideas would be clarified dramatically in the next generation, and now provide a confident starting point for the physicists' approach to ever more complex systems, far from where Keith and his colleagues began.

Although more would come, this early work on the many--body problem was the primary motivation for the honors bestowed on Keith over the years:  the call to an endowed chair at the University of Pennsylvania (1956); the Dannie Heineman Prize for Mathematical Physics from the American Physical Society (1963); 
election to the American Academy of Arts and Sciences (1968) and the National Academy of Sciences (1969); and an Honorary Doctorate from Indiana University (1976).

\section{UCSD}

\begin{quote}
In the late fall of 1958 I spent  a few days in San Diego consulting at General Atomic.  Two geophysicists, Leonard Lieberman and Carl Eckart, at the Scripps Institute of Oceanography, knew that I was visiting  ... they asked me to have lunch with them the next day ... When they arrived to pick me up at General Atomic, they had brought along a very tall suntanned man, Roger Revelle.\footnote{KAB 15 Feb 1994. From the UCSD archives.}
\end{quote}

Roger Revelle was the director of the Scripps Institute of Oceanography, and had been given the task of planning a new campus for the University of California.  What today is the University of California, San Diego (UCSD) was then largely vacant land in La Jolla, partially occupied by Camp Matthews, a Marine training base.   It was (and is) a spectacular location.  It also was the edge of civilization.  As Revelle noted in his remarks at Keith's retirement dinner, not everyone could see the great University that would rise in that empty space.\footnote{There were more concrete problems, since much of La Jolla residential real estate was then governed by restrictive covenants that prevented sales to Jews and others.  These would not be fully removed until 1968, with passage of the Fair Housing Act.} 

Keith agreed to join the adventure, along with James Arnold and Harold Urey in Chemistry and David Bonner in Biology.   His immediate responsibility was to build a physics department, but  he would influence almost every aspect of academic life on the new campus.  In his description, it was an almost magical time of generous and enthusiastic state support for higher education.

Keith moved to La Jolla in the Fall of 1959.  By 1962, the physics department included theorists  Walter Kohn, Norman Kroll, Maria Goeppert Mayer, Marshall Rosenbluth, and Harry Suhl, as well as experimentalists George Feher and Bernd Matthias, all of whom would be elected to the National Academy of Sciences; also in the first group of recruits were  Geoffrey and Margaret Burbidge, who would become Fellows of the Royal Society of London.   The Burbidges, still fresh from their pioneering work on the origin of the elements, provided the nucleus for an astrophysics group, and Margaret also would  have an extraordinary impact on the status of women in physics and astronomy, both by example and through explicit advocacy.  Kohn did his foundational work on density functional theory, recognized by the 1998 Nobel Prize in Chemistry, while on the UCSD faculty.  Feher would bring his spectroscopic talents to bear on the problems of photosynthesis, launching a whole field of biological physics.  Plasma physics and condensed matter were (then) not so well represented at most US universities, and the first cohorts of PhD students at UCSD included many future leaders in these fields.  Keith exhibited good taste;  with Revelle and the resources of the University of California behind him, he was also very persuasive.

Keith was especially proud of bringing  Maria Goeppert Mayer to UCSD in 1960. Mayer would share the 1963 Nobel Prize in Physics for her theoretical work on the nuclear shell model, but her career trajectory meandered through a minefield of misogyny. Keith was delighted that the University of California could do the right thing and appoint her Professor of Physics, with no asterisks on account of her gender or marital status;\footnote{In those years, policies with the stated goal of combating nepotism often were cited by universities as reasons not to appoint the female half of even the most distinguished academic couples.} coincidentally the Chemistry Department could also add a distinguished theorist to its  ranks (Joseph Mayer).   To capture the environment in which these decisions were being made, it is worth recalling the local newspaper headline:  ``San Diego Mother Wins Nobel Physics Prize.''

In addition to being the founding chair of the Department of Physics, Keith recruited leaders for  the Department of Mathematics and the engineering school.  He helped  bring some of the first literary scholars  and linguists to the campus, as well as George and Jean Mandler, who formed the nucleus of a very forward looking Department of Psychology.  As he so often did in those early days, Keith flew out to see the Mandlers in Toronto; George recalled\footnote{G Mandler, {\em Interesting Times. An Encounter with the 20th Century, 1924--} p 182 (Psychology Press, New York, 2013).} ``... [a] day spent sitting on the floor in our living room, hearing about the promise of UCSD ... by the time Keith left I was sold.''  In philosophy he recruited the prolific Avrum Stroll, who would become a dear friend, and who in turn recruited Herbert Marcuse.

In those early years, Keith  served on the ``Director's Administrative Advisory Council," on committees to oversee  growth of the humanities and social sciences, the construction of almost every building, planning  for undergraduate education, the establishment of  residential colleges, and recruitment of faculty in the fine arts.  He chaired committees to review the campus master plan, and to look ahead to a second decade. He brought the first computer  and  recruited the first librarian. All of this in just the few years 1959--65.

Keith's success in recruiting  became the stuff of legend.  He enjoyed recounting the process, describing his constant travel and discussions, how he used each positive indication from one candidate as part of the sales job for the next.  The first recruits were united by a spirit of adventure which stayed with them for decades, and many saw Keith as the most adventurous of all.

The real history  is messier than the legends.  The whole process of creating UCSD began very informally.  Faculty were being recruited to a university that did not quite exist, with no departments or department chairs.  Walter Kohn  insisted that he would come only if Keith were appointed as chair of the physics department once things solidified.  As the structures began to form, Keith was hugely disappointed that Roger Revelle was not asked to serve as the new campus's first chancellor.    This disappointment certainly played a role in his decision to take a leave of absence from UCSD, serving as vice--president of the Institute for Defense Analyses (see below).  When Keith returned to campus, he was appointed Dean, but correspondence from that time reveals surprising ambiguity about his actual responsibilities.

By the mid--1960s, conflicts with the upper administration became more frequent and Keith moved away from his multiple responsibilities.  One late contribution was a vision for continued growth of the Physics Department he had founded.\footnote{KAB 22 Oct 1965.  From the UCSD archives.} He argued that the Department needed to grow by roughly a factor of two from 1965 to 1972, and that this growth needed to be balanced across many different fields of physics in order to meet the emerging intellectual challenges.  Particularly noteworthy was his proposal for the growth of biophysics as a branch of physics, and the idea that this should be supported in part with resources drawn from the medical school.  This was in clear opposition to some of his prominent biologist colleagues, who saw biophysics as a branch of biology and argued against any physics--centered efforts.  Keith's argument won out, and UCSD became the first major US physics department to have a significant presence in the field, which continues today.

By any measure, the project of building UCSD was a great success.  Keith Brueckner and Roger Revelle probably contributed more to this success than any other individuals. Nonetheless, in private discussions I sometimes sensed that Keith was a little disappointed.  What he and his colleagues had built became a great university, but one rather like all the other great universities of its time.  He, and they, had hoped for something different.  He foresaw the problems of teaching in large lecture halls rather than in intimate discussions at the blackboard, and stayed long enough the see the start of declining state support for public universities, as well as political meddling in university governance. But his enthusiasm for physics, for science more generally, and for the special intellectual life of the university, persisted until the end.

\section{Advice and dissent}
\label{jason}

As with many theoretical physicists of his generation,  Keith spent time at Los Alamos  consulting on the nuclear weapons program and taking advantage of the computing facilities to push forward his basic scientific work.   In the late 1950s, a handful of those colleagues began to think that additional structure was needed in order to provide the government with the best possible scientific advice.   A first step in this direction was Project 137, a summer study group led by John Wheeler which included Keith, Murph Goldberger, and Kenneth Watson, among others.  These three would take the lead in a more ambitious effort that was organized under the auspices of the Institute for Defense Analyses (IDA).  Together with Murray Gell--Mann and Charles Townes (then vice--president of IDA), they formed the steering committee for a consulting group called JASON.\footnote{Not an acronym, but a playful suggestion by Mildred Goldberger.}

The initial JASON group included Subrahamanyan Chandrasekhar, Freeman Dyson, Val Fitch, Donald Glaser, Norman Kroll, Leon Lederman, Francis Low, Walter Munk, Malvin Ruderman, Edward Salpeter, and Sam Treiman.  There was also an effort to engage promising younger people, including Henry Kendall and Steven Weinberg.  An important part of the plan was that the group would not be asked specific questions by government agencies, but rather be presented with a variety of issues and concerns out of which they would formulate questions for deeper investigation.  This independence, part of the original vision articulated by Keith and his colleagues, would eventually become a source of friction between JASON and its government sponsors.

In 1961, Keith took a leave from UCSD to serve as vice--president of the Institute for Defense Analyses, taking over from Townes.  For a bit more than a year he helped   build the IDA technical staff, with much the same energy that he had brought to recruiting at UCSD.  But as Keith himself described it,\footnote{KAB Feb 15, 1994.  From the UCSD archives.}   ``In my enthusiasm and inexperience with the Washington scene, I had some conflict with the governing board of IDA ....''    He was also  disappointed by the appointment of Richard Bissell as president of IDA; Bissell had been deputy director of the Central Intelligence Agency, where he led the planning for the Bay of Pigs invasion.  Keith returned to UCSD in 1963, and stepped away from JASON.  He continued to consult, but never again took a leadership position.\footnote{In 1983 I had been offered the opportunity to spend time at one of the weapons labs.  Keith's advice was simple:  if you can possibly do without the money, don't go.  ``You have nothing more valuable,'' he said, ``than the time to think about the problems you find important.''}

While still at IDA, Keith took time to lecture at Georgetown.\footnote{D Breasted, Physicist says students are too wary of science.  {\em The Evening Star}  February 15, 1962, p. C--14 (Washington DC).} He emphasized that massive support for military research was having a detrimental effect on science more broadly, pulling people away from fundamental problems.  More subtly, he argued that the scale of funding was distorting the marketplace and effectively reducing the attractiveness of teaching in high schools and small colleges.  These observations seem quite prescient.  He also railed against the image of scientists as geniuses, and the mistaken impression that science is too difficult.

After Keith's departure from the group, JASON was subject to increasing criticism for its involvement with the Vietnam War.   By May 1970, in the wake of escalation into Cambodia, UCSD buildings  (as on many campuses) were occupied and Keith was targeted by students aware of his continued engagement with the defense establishment.  He responded in the campus newspaper, stating his positions on the issues of the day:  (1) The United States should withdraw immediately from Southeast Asia.  (2) The federal government should transfer research funds from the defense agencies to the National Science Foundation. (3) The University of California should terminate management of the nuclear weapons laboratories.  (4) The rapid growth of science at UCSD led to ``unbalanced structure, detrimentally affecting the over--all quality ... retarding the growth in humanities, social sciences, and fine arts.''  The next year Keith and his wife Elsa hosted an anti--war fund raising event featuring the recently paroled activist David Harris, along with actors Jane Fonda and Donald Sutherland.

Keith believed strongly that policymakers, in particular those working on defense and national security, needed the best possible scientific advice.  He held to this position despite his strong opposition to the policies that were being made.    Although his version was especially stark, many physicists of his generation found themselves similarly caught between advice and dissent.

\section{Controlled fusion}
\label{fusion}

\begin{quote}
... summer of 1955.  Ken Watson recruited Keith Brueckner, Geoff Chew, Francis [Low], and me to work on what was then the classified program on controlled fusion.  This was a rather bizarre experience. ... somehow Keith had become an expert on plasma physics.  Geoff, Francis, and I didn't know beans ... .\footnote{ML Goldberger. In {\em Asymptotic Realms of Physics. Essays in Honor of Francis E.~Low}, AH Guth, K Huang, and RL Jaffe, eds, pp xi--xv (MIT Press, Cambridge MA, 1983).}
\end{quote}

Keith's old friend Murph Goldberger reminds us here that the search for fusion power, which today is the subject of frequent news articles, started as a secret project.  The dream was to harness the energy source that powers the sun for use here on earth. 

The initial and still dominant idea was to use magnetic fields to confine a plasma, which could  then reach the extreme temperatures and densities needed for fusion to occur at high rates.  The invention and rapid development of optical lasers in the early 1960s led many people to wonder if a powerful pulse of laser light could be used to send controlled shock waves through a material target.  In his role at the Institute for Defense Analyses, Keith organized a summer study on high powered lasers in 1963, and this was one of the first places where the possibility of laser fusion was discussed.

It was clear that progress  would require a combination of theory, simulation, and rather large scale experiments.  In particular, a wide variety of unstable flows can disrupt the extreme conditions that are the goals both in magnetic confinement and in laser driven fusion.  Keith was fascinated by the challenging theoretical problems, and he was no stranger to large computations.  He and his colleagues seem to have been near the origin of the idea to use multiple lasers as the hydrodynamic driver for direct implosion of a spherically symmetric composite target.  This work in 1969--70 was classified by the Atomic Energy Commission (AEC), and would be declassified over a period of  several years, culminating in publication of a major technical review article written with Siebe Jorna in 1974. Even in 1975, Keith could not tell how much of his path through the subject had been charted independently by colleagues at the AEC laboratories.\footnote{KAB 3 Feb 1975.  From the UCSD archives.}

Much of Keith's foundational work on direct drive laser fusion was done while on leave from UCSD to KMS Fusion, a private company.    Based on this theoretical work, and after much negotiation, KMS obtained a no--cost contract from the AEC to continue from 1971.  Keith became the Technical Director of the company and (once again) recruited a substantial staff.    By 1974 there was a working experimental system that was performing diagnostics to test the theory and achieving neutron--producing implosions as multiple laser pulses converged symmetrically on spherical targets with glass shells.   Keith stepped away from KMS shortly after, and  experimental work continued under the direction of Robert Hofstadter.

Keith continued to work on physics  relevant to laser fusion after returning to UCSD.  At the start, he had to produce written versions of his seminars, knowing an FBI agent would be following along to be sure he didn't deviate from the script.   He was commissioned by the Electric Power Research Institute to assess the prospects for laser fusion, and assembled a team of colleagues.  Their impactful report, published in 1977, included fascinating windows into previously classified data.

Keith returned to Les Houches to lecture at the 1980 Summer School on laser--plasma interactions, sharing his excitement with a new generation.\footnote{He would come back to Les Houches once more, just for a visit, in 2001.  It was a pleasure to hear him reminiscing,  both about physics and about the evolution from the spartan facilities of 1958 to the comfort of the modern accommodations.} He took a special interest in the progress of fusion research in the Soviet Union, and had a correspondence with Nikolay Basov.  They had hoped to meet at a conference in La Jolla in 1980, but US visas for Soviet scientists were revoked in response to the internal exile of Andrei Sakharov.  Closing the circle, Nikolay's son Dmitri Basov would join the UCSD physics department a few years after Keith's retirement, and was Chair at the time of Keith's death.

\section{Mountains and family}

Keith had a lifelong passion for climbing.  He made regular visits to the Alps, engaging in very technical climbs, often with guides.  With time he also came to enjoy the good food and wine available nearby, and would amass a serious cellar of his own.  He was a member of the American Alpine Club, and in some versions of his curriculum vitae this appears just below his membership in the National Academy of Sciences.  When traveling without his family he would recount his adventures in letters and postcards, sometimes in more detail than necessary.  As with his accounts of different adventures in physics, Keith could tell a story without needing to be the hero, and there is nothing heroic about waiting out two days of rain while trapped in a tent with your guide.

\begin{figure}
\includegraphics[width=\linewidth]{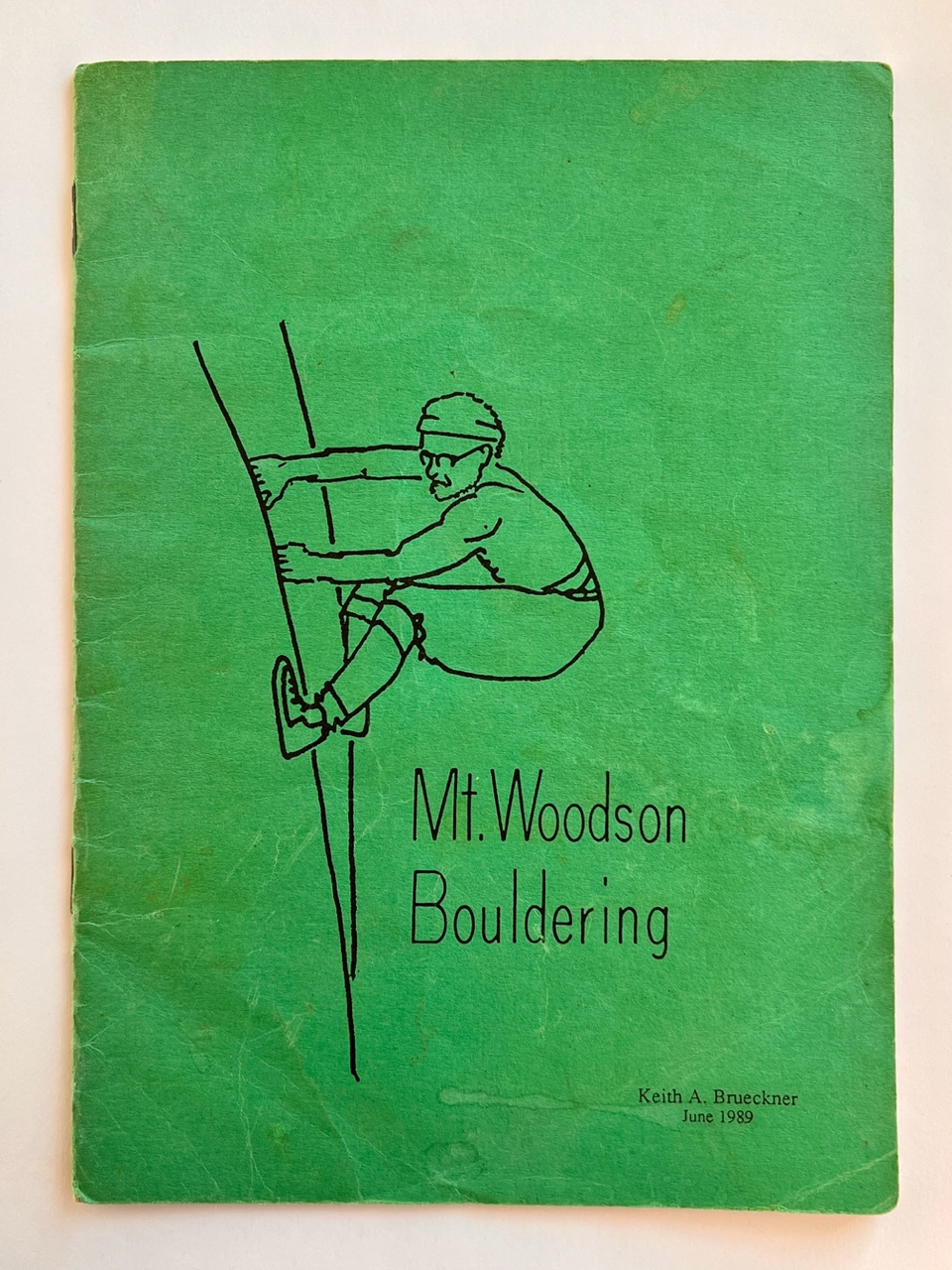}
\caption{This guide circulated widely in the Southern California climbing community.  Photo by CD Bialek. \label{woodson}}
\end{figure}

Closer to home, Keith climbed regularly at the boulders in Joshua Tree National Park and at Mount Woodson.  He often introduced young colleagues and students to technical climbing, and with time he enjoyed a reputation as the old man of the mountain.  He literally wrote the book on his favorite climbs at Mount Woodson  (Fig \ref{woodson}), and copies of this pamphlet circulated widely in the Southern California climbing community.  A fuller account of his climbing career, capturing the charm of his letters, was published posthumously.

As he finished his brief postwar masters degree in Minnesota, Keith married Marjorie Dumas and they moved together to Berkeley.  Their two sons, Jan and Anthony (Tony),  have had distinguished academic careers of their own.  Jan is an economist, now at the University of California, Irvine, and Tony was a philosopher at the University of California, Santa Barbara.   Marjorie was Keith's partner through the years at Indiana University and the University of Pennsylvania, when he did his most impactful physics.  They separated shortly after their move to La Jolla, and eventually Marjorie returned to work as an artist.  Fittingly, one of her paintings adorns the cover of Tony's last book.

By the time Keith married Elsa Dekking, she had four young daughters---Charlotte, Barbara, Jessica, and Carolyn---and they would have a fifth, Leslie, together;  the five girls grew up with `Keithie' as their dad.  Born in the Netherlands and arriving in La Jolla via Venezuela, Elsa was Keith's  partner in the human side of the recruiting effort that built UCSD.  Their older daughters  have had diverse professional lives in the arts, education, farming, animal science, and graphic design; their youngest is a public interest lawyer who has argued before the US Supreme Court.  Keith and Elsa separated in 1981.

Bonnie Lichtenstein married Keith in 1988.  She worked as a security administrator at Physical Dynamics, Lockheed, United Technology, and finally the Institute for Defense Analyses, and so had strong connections to an important chapter of Keith's earlier life.  They lived  in La Jolla,  enjoyed frequent travels, and were close to Bonnie's children Deborah and Patrick.    In his final years, Keith suffered with an aggressive dementia, and Bonnie shouldered this burden with courage and grace. He died on 19 September 2014.

Finally, in appreciation (and in full disclosure), I should add that Keith was my father--in--law.  I have a vivid memory of my first encounter with him, when Charlotte and I went to visit just a few months after we had met.  The physics questions began as he was driving us to dinner, and continued through the meal.  It was like revisiting my qualifying exam, but with higher stakes.  It seems I passed, and  talking physics with Keith became a continuing pleasure.  He also provided a more personal connection to the generation of my mentors.  Keith was not an easy personality, and on learning that I was his son--in--law my senior colleagues often replied with amusing stories.  

Keith spent his final years in a memory care facility, a place he described as ``an elegant jail.''   In the garden, on a beautiful afternoon, we reminisced, but soon the conversation became serious.  We spoke, for one last time, about  what was new and exciting in physics.

\begin{acknowledgments}
I am grateful for input from the family: Charlotte Bialek, Jan Brueckner, Leslie Brueckner, and Carolyn Miller; and from the late George Mandler. 
My sincere thanks also to Colleen Garcia and her colleagues at the UCSD Archives; to Alexandra Briseno at the NASEM Archives; and to Gordon Baym, Jill Dahlburg,  and David Zierler for their comments on earlier drafts.
\end{acknowledgments}

\section*{Selected papers}

\begin{description}
\item[1949] With ML Goldberger. The excess of negative over positive mesons produced by high energy photons. {\em Physical Review} {\bf 76,} 1725.
\item[1950] The production of mesons by photons. {\em Physical Review} {\bf 79,} 641--650.
\item[1952]  Meson--nucleon scattering and nucleon isobars.  {\em Physical Review} {\bf 86,} 106--109.
\item[1955] With CA Levinson. Approximate reduction of the many--body problem for strongly
interacting particles to a problem of self--consistent fields.  {\em Physical Review}  {\bf 97,} 1344--1352.\\
\\
Two--body forces and nuclear saturation. III.  Details of the structure of the nucleus. 
{\em Physical Review} {\bf 97,} 1353--1366. \\
\\
Many--body problem for strongly interacting particles. II. Linked cluster expansion. 
 {\em Physical Review} {\bf 100,} 36--45. 
\item[1956] With W Wada. Nuclear saturation and two--body forces: Self--consistent solutions and the effects of the exclusion principle.   {\em Physical Review} {\bf 103,} 1008--1016.
\item[1957] With M Gell--Mann. Correlation energy of an electron gas at high density.  {\em Physical Review} {\bf 106,} 364--368.
\item[1958] With JL Gammel. Properties of nuclear matter.   {\em Physical Review} {\bf 109,} 1023--1039.
\item[1959] Theory of nuclear structure, and Further applications of the theory of many--body systems. In {\em The Many Body Problem. Le Probl\`eme \`a N Corps. Cours donn\'es \`a l'Ecole d'\'Et\'e Physique Th\'eorique, Les Houches, Session 1958}, C DeWitt and P Nozieres, eds, pp 47--241 (John Wiley and Sons, New York).
\item[1968] With S--K Ma.  Correlation energy of an electron gas with a slowly varying high density. {\em Physical Review} {\bf 165,} 18--31.
\item[1974] With S Jorna.  Laser--driven fusion. {\em Reviews of Modern Physics} {\bf 46,}   325--367.
\item[1977] With RW Gross, GR Hopkins, S Jorna, RE Kidder, GL Kulcinski, JH McNally, and WB Thompson. An assessment of laser--driven fusion. {\em International Journal of Fusion Energy} {\bf 1,} 25--54.
\item[1982] Principles of laser confinement fusion.  In {\em Laser--Plasma Interaction,  Proceedings of the 34th Summer School on Theoretical Physics, Les Houches.} R Balian and JC Adam, eds, pp 539--644 (North--Holland, Amsterdam).
\item[1989] {\em Mt.~Woodson Bouldering.}   Self--published (La Jolla CA)
\item[2016] {\em Mountaineering: A Personal History.}  (Archway Publishing, Bloomington IN).
\end{description}

\end{document}